\documentclass[submission, Phys]{SciPost}
\usepackage{braket}

\newcommand\mc[1]{\mathcal{#1}}
\newcommand\mn[1]{\text{#1}}
\newcommand\nk[1]{{\left( #1 \right)}}
\newcommand\ek[1]{{\left[ #1 \right]}}
\newcommand\gk[1]{{\left\lbrace #1 \right\rbrace}}

\usepackage{xcolor}

\hypersetup{
    colorlinks,
    linkcolor={red!50!black},
    citecolor={blue!50!black},
    urlcolor={blue!80!black}
}

\usepackage[bitstream-charter]{mathdesign}
\DeclareSymbolFont{usualmathcal}{OMS}{cmsy}{m}{n}
\DeclareSymbolFontAlphabet{\mathcal}{usualmathcal}

\begin{document}

\begin{center}{\Large \textbf{
Aberration of the Green's function estimator in hybridization expansion continuous-time quantum Monte~Carlo
}}\end{center}

% TODO: write the author list here. Use first name (+ other initials) + surname format.
% Separate subsequent authors by a comma, omit comma and use "and" for the last author.
% Mark the corresponding author with a superscript star.
\begin{center}
Andreas Hausoel\textsuperscript{1,2$\star$},
Markus Wallerberger\textsuperscript{3},
Josef Kaufmann\textsuperscript{3},\\
Karsten Held\textsuperscript{3}
and Giorgio Sangiovanni\textsuperscript{2},
\end{center}

\begin{center}
{\bf 1} Institute for Theoretical Solid State Physics, Leibniz IFW Dresden,\\Helmholtzstr. 20, 01069 Dresden, Germany
\\
{\bf 2} Institut f\"ur Theoretische Physik und Astrophysik\\
and W\"urzburg-Dresden Cluster of Excellence ct.qmat,\\
Universit\"at W\"urzburg, 97074 W\"urzburg, Germany
\\
{\bf 3} Institute of Solid State Physics, TU Wien, 1040 Vienna, Austria
\\
${}^\star$ {\small \sf a.hausoel@ifw-dresden.de}
\end{center}

\begin{center}
\today
\end{center}

% For convenience during refereeing (optional),
% you can turn on line numbers by uncommenting the next line:
%\linenumbers

\section*{Abstract}
{\bf
We describe an aberration of the resampling estimator for the Green's function customarily used in
hybridization expansion continuous-time quantum Monte Carlo.
It occurs due to Pauli principle constraints
in calculations of
Anderson impurity models
with baths consisting of a discrete energy spectrum.
We identify the missing Feynman diagrams, characterize the affected models and discuss implications as well as solutions.
This issue does not occur when using worm sampling or in the presence of continuous baths. 
However certain energy spectra can be inherently close to a discrete limit, 
and we explain why autocorrelation times can become very large in these cases.
}

\vspace{10pt}
\noindent\rule{\textwidth}{1pt}
\tableofcontents\thispagestyle{fancy}
\noindent\rule{\textwidth}{1pt}
\vspace{10pt}

\section{Introduction}
\label{sec:intro}
% TODO: write your article here.
Despite decades of intense research, a generic solution to
the quantum many-body problem is still lacking.
The use of diagrammatic Monte Carlo techniques, however, has
led to significant progress for special cases of interest,
such as for the Anderson impurity model (AIM) and its
generalizations to correlated molecules and retarded
interactions \cite{gull2011continuous}.

Diagrammatic Monte Carlo techniques
proceed in a three-step fashion:
firstly, the
action is split into two parts,
where one part is solved exactly.
Secondly, the other part of the action is treated by
expanding
an appropriate generating function
(such as the partition function
or free energy)
with respect to it.
Thirdly and finally, the resulting probability distribution is sampled using
Markov chain Monte Carlo.

Unlike the classical case, in quantum mechanics each observable
is operator-valued and comes with its own generating function. Thus after a series expansion, it
has its own probability distribution to be sampled.
In principle,
Monte Carlo algorithms
have to sample
these distributions separately,
e.g., the mean density, the one-particle propagator, and even
the propagator evaluated at different times or orbitals.

Two methods are known to deal with this issue: (i) worm sampling \cite{Prokofev_hfye,Prokofev_ct},
where one forms the direct sum over the probability spaces of all observables considered,
and samples that compound distribution.
This quickly leads to an unwieldy number of computations.
If the distributions of interest are similar in their structure,
which is often the case,
with (ii) resampling\footnote{This is commonly employed for the diagrammatic Monte Carlo calculation of the Green's function in the AIM\cite{Rubtsov,werner-continuous-time-2006}.} one can sample only a single distribution and
map all other observables to different estimators with respect to
that distribution. Resampling is algorithmically simpler but yields an incomplete estimator
(and thus wrong results) if the mapping is not surjective. One also runs
into autocorrelation problems if the mapping is indeed surjective, but the
probability distributions are substantially different.

A widely used,
state-of-the-art finite-temperature diagrammatic solver for Anderson impurity
models, is continuous-time quantum Monte Carlo
in the hybridization expansion\cite{werner-continuous-time-2006} (CTHYB),
where the partition function is expanded with respect to the bath hybridization.
One usually employs resampling for measuring the Green's function, relating
each Green's function diagram to a process of ``cutting'' parts off a
diagram in partition function ($Z$) space. This is already known to fail
for equal-time correlators,
certain higher order Green's functions, and close to
the atomic limit. There, worm sampling must be used instead \cite{wormprb}.
However, resampling is still widely used because it is believed
to succeed away from the aforementioned cases.

In this paper, 
we identify one more incompleteness
of the resampling $G$-estimator
for certain finite systems, curtailing the
viability of the method in quantum chemistry applications. 
In analogy to optics
we call the phenomenon \emph{aberration},
which means an image being blurred or distorted.
Let us note, that the essence of the problem has been 
already described in one of the authors' thesis \cite{hausoel2022electronic},
and noticed
independently of us \cite{Melnick} recently.
Here, we further
determine all effected systems and
show that for certain infinite systems,
this form of resampling, while formally consistent, causes the
autocorrelation length to grow significantly.
Moreover, we point a way out of this problem, by using worm sampling.

The paper is organized as follows:
In Section~\ref{s_cthyb} we give a
short review of the CTHYB algorithm,
before we identify missing Feynman diagrams
in systems with finite bath size
in Section~\ref{s_incos}.
In Section~\ref{s_monomer1}
we show an example and make
the link to autocorrelation times.
Finally in Section~\ref{s_inf_sys},
we show a system with infinite
bath size and autocorrelation problems,
before we conclude in Section~\ref{sec:conclusion}.

\section{CTHYB and the measurement of the Green's function}
\label{s_cthyb}
In this section we repeat the basic concepts of CTHYB,
namely the expansion formulas and
how they relate to Feynman diagrams,
as well as the differences between the two types of measuring the
Green's function:
$Z$-sampling, corresponding to (ii) resampling, and $G$-sampling, corresponding to (i) worm sampling.
For more details we refer the reader to the corresponding literature\footnote{
Our notation is based on chapter II of Reference~\citenum{wormprb}
and chapter II of Reference~\citenum{sssprb}.
For all details take a look at Reference~\citenum{gull2011continuous}.}.

\subsection{The expansion formulas and Feynman diagrams}
The Hamiltonian of the multi-orbital AIM reads
\begin{align}
  \hat{H}_{\mn{AIM}} &= \hat{H}_{\mn{bath}} + \hat{H}_{\mn{hyb}}^\dagger + \hat{H}_{\mn{hyb}}+ \hat{H}_{\mn{loc}}
    \label{eq_aim_hamiltonian} \\
    &= \sum_{{p}\mu} \epsilon_{{p\mu}} \hat{a}_{{p}\mu}^\dagger \hat{a}_{{p}\mu}
    + \sum_{{p}\mu\nu}  V_{{p}\mu\nu}^* \hat{a}_{{p}\mu}^\dagger \hat{c}_\nu +
      \sum_{{p}\mu\nu}  V_{{p}\mu\nu} \hat{c}_\nu^\dagger \hat{a}_{{p}\mu}  \nonumber \\
    & \phantom{=} + \hat{H}_{\mn{loc}} [\hat{c}^\dagger,\hat{c}] \nonumber.
\end{align}
Operators $\hat{c}^\dagger_\nu$ ($\hat{c}_\nu$)
create (annihilate) electrons on the impurity with flavor $\nu$,
whereas operators $\hat{a}^\dagger_{p\mu}$ ($\hat{a}_{p\mu}$)
create (annihilate) electrons on the $p$-th bath site,
which has an energy of $\epsilon_{{p\mu}}$
and belongs to impurity flavor $\mu$.
In the second and third term of Eq.~\eqref{eq_aim_hamiltonian},
each impurity flavor $\nu$
couples to its own
non-interacting bath sites
with amplitudes $V_{p\nu\nu}$
(diagonal hybridization),
but may also couple to the bath sites
of other impurity flavors $\mu$ via $V_{p\mu\nu}$
with $\mu\neq\nu$ (off-diagonal hybridization).
The fourth term
$\hat{H}_{\mn{loc}} [\hat{c}^\dagger,\hat{c}] = -t_{\mu\nu}\hat{c}_\mu^\dagger \hat{c}_\nu +
U_{\kappa\lambda\mu\nu} \hat{c}^\dagger_\kappa \hat{c}^\dagger_\lambda \hat{c}_\nu \hat{c}_\mu $
contains the one-
and two-particle
interaction on the impurity.

The CTHYB expansion of the partition function is \cite{gull2011continuous, sssprb}
\begin{align}
    Z&= \sum_{k=0}^\infty \frac{1}{k!}
    \int \mn{d}^k \mc{C} \; \int \mn{d}^k \mc{C}' \;
    w_{\mn{loc}}(\mc{C},\mc{C}')
    w_{\mn{bath}}(\mc{C},\mc{C}'),
    \label{eq:Z}
\end{align}
i.e., an integral over configurations $(\mc{C},\mc{C}')$
of an appropriate weight function.
More specifically, one defines a configuration
$\mc{C}=\gk{(\nu_1,\tau_1), \dots, (\nu_k,\tau_k)}$
as a set of times and flavors for the creation operators,
and
$\mc{C}'=\gk{(\nu_1',\tau_1'), \dots, (\nu_k',\tau_k')}$
as the corresponding set for the annihilation operators,
where $k$ is the expansion order.
The integral in Eq.~(\ref{eq:Z}) thus is
\begin{equation}
    \int \mn{d}^k\mc{C} \equiv \sum_{\nu_1} \dots \sum_{\nu_k}
    \int_0^\beta \mn{d}\tau_1
    \int_0^\beta \mn{d}\tau_2
    \dots
    \int_0^\beta \mn{d}\tau_{k}
\end{equation}
and the local weight is given by
\begin{align}
  w_{\mn{loc}} (\mc{C}, \mc{C}') &=
  \mn{Tr}_c \ek{  \mn{e}^{-\beta \hat{H}_{\mn{loc}}}
          T_\tau \prod_{i=1}^k \hat{c}_{\nu_i}^\dagger(\tau_i)
                   \hat{c}_{\nu_i'}(\tau_i')
  }.
\end{align}
Here, the trace $\mn{Tr}_c [\dots] = \sum_s \bra{s} \dots \ket{s}$
is computed over a complete many-body basis
of the impurity.
For two flavors (one spin-1/2 band) such a basis would be
$\ket{s} \in \gk{ \ket{0}, \ket{\uparrow}, \ket{\downarrow},\ket{\uparrow\downarrow}}$.
The argument of the trace is a time-ordered product
of impurity operators,
whose time-evolution is governed by $\hat{H}_{\mn{loc}}$
via $\hat{c}_{\nu} (\tau) = \mn{e}^{\hat{H}_\mn{loc}\tau} \hat{c}_{\nu} \mn{e}^{-\hat{H}_\mn{loc}\tau}$.

The bath weight
describes the retardation effect of the bath on the impurity
and is given by
\begin{align}
    w_\mn{bath} (\mc{C},\mc{C}')
  &= \mn{det}
  \begin{pmatrix}
      \Delta_{\nu_1\nu_1'}(\tau_1\!-\!\tau_1') & \! \dots\! & \Delta_{\nu_1,\nu_k'}(\tau_1\!-\!\tau_k') \\
      \vdots & \ddots & \vdots \\
      \Delta_{\nu_k\nu_1'}(\tau_k\!-\!\tau_1') &\! \dots \!& \Delta_{\nu_k,\nu_k'}(\tau_k\!-\!\tau_k')
  \end{pmatrix},
\end{align}
where the matrix elements are $\Delta_{ij} = \Delta_{\nu_i,\nu_j'} (\tau_i - \tau_j') $
with $(\nu_i,\tau_i) \in \mc{C}$ and
$(\nu_j',\tau_j') \in \mc{C}'$.
The propagator is the hybridization function \cite{gull2011continuous}
\begin{equation}
  \Delta_{\nu\nu'}(\tau) = \sum_{p=1}^{N_p} \frac{V_{p\mu\nu} V_{p\mu\nu'}^*}{\mn{e}^{\beta\epsilon_{p\mu}}+1} \times
  \begin{cases}
    \mn{e}^{\epsilon_{p\mu} \tau}, & \mn{if}\ \tau > 0, \\
    -\mn{e}^{\epsilon_{p\mu} (\beta-\tau)}, & \mn{if}\ \tau < 0,
  \end{cases}
\end{equation}
which is a sum of the non-interacting Green's functions
of the $N_p$ bath sites,
weighted with the hopping amplitudes
$V_{p\mu\nu} V_{p\mu\nu'}^*$
($N_p$ can in principle be infinite).
Therefore it contains three processes combined:
the hopping from impurity to bath,
propagation through the bath,
and hopping back from bath to impurity.

\begin{figure}[tp]
    \centering
    %\captionsetup{justification=raggedright}
    \includegraphics[width=0.4\linewidth]{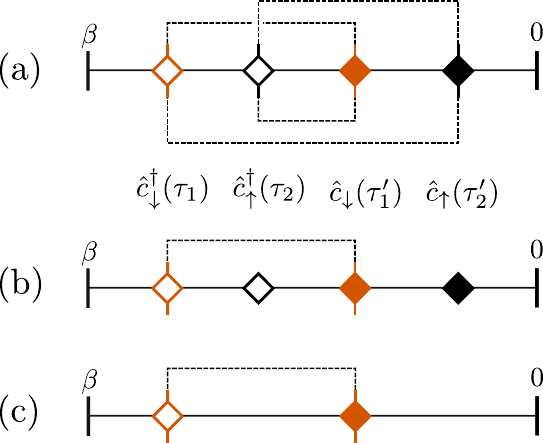}
    \caption{
      (a), (c) $Z$-configurations for an AIM with one orbital.
      Empty diamonds are creation operators,
      filled diamonds annihilation operators,
      and color denotes impurity flavor.
      The dashed lines mean a propagation of the electron through the bath.
      (b) Derived Green's function configuration obtained either (i) from (c) through worm sampling inserting two (black) worm operators at
      $\tau_2$ and $\tau_2'$, or (ii) from (a) through $Z$-sampling removing the hybridization lines from the black operators.
 }
  \label{diagramme_monomer1_intro}
\end{figure}

Let us take the
partition function configuration
$\hat{c}_\downarrow^\dagger(\tau_1) \hat{c}_\uparrow^\dagger(\tau_2) \hat{c}_\downarrow(\tau_1') \hat{c}_\uparrow(\tau_2') $
of an AIM with two flavors $\uparrow$ and $\downarrow$
as example of a $w_{\mn{bath}}$, see Fig.~\ref{diagramme_monomer1_intro} (a).
This corresponds to
$\mc{C} = \gk{ (\downarrow,\tau_1), (\uparrow,\tau_2)}$ and
$\mc{C}' = \gk{ (\downarrow,\tau_1'), (\uparrow,\tau_2')}$
in the abbreviated notation.
The filled (empty) diamonds indicate the annihilation (creation) of an electron
on the impurity.
The dashed lines attached to an operator represent a connection
of this operator to the bath.
In our example a pair of an impurity electron and bath hole
is created at $\tau_2$.
The bath hole
propagates to $\tau_2'$,
where it annihilates with another impurity electron;
the propagator for these three processes combined
is $\Delta_{\uparrow\uparrow}(\tau_2-\tau_2')$.
The same happens with another hole from $\tau_1$ to $\tau_1'$ via
$\Delta_{\downarrow\downarrow}(\tau_1-\tau_1')$.
A second possibility is that the hole at $\tau_2$ propagates to $\tau_1'$
via $\Delta_{\uparrow\downarrow}(\tau_2-\tau_1')$,
and the other hole from $\tau_1$ to $\tau_2'$
via $\Delta_{\downarrow\uparrow}(\tau_1-\tau_2')$.
All these propagations
can compactly be written as a determinant
\begin{align}
  w_\mn{bath} (\mc{C}, \mc{C}')
  &= \mn{det}
  \begin{pmatrix}
    \Delta_{\downarrow\downarrow}(\tau_1-\tau_1') & \Delta_{\downarrow\uparrow}(\tau_1-\tau_2') \\
    \Delta_{\uparrow\downarrow}(\tau_2-\tau_1') & \Delta_{\uparrow\uparrow}(\tau_2-\tau_2')
  \end{pmatrix},
\end{align}
which corresponds to the application of Wick's theorem
to the non-interacting bath.

\subsection{Resampling and worm sampling}
Now we discuss the two types of measuring the Green's function.
The hybridization expansion of the Green's function is
\begin{align}
  G_{\nu\nu'} (\tau-\tau') &= \frac{1}{Z} \sum_{\mc{C}}
  w_{\mn{loc}}(\mc{C}\cup \{(\nu,\tau)\}, \mc{C}'\cup \{(\nu',\tau')\} )
   \nonumber \\
  &\times w_{\mn{bath}}(\mc{C},\mc{C}'). \label{eq_g_expansion}
\end{align}

(i) Worm sampling takes a partition function configuration $(\mc{C},\mc{C}')$,
which was generated by the Markov chain,
and adds $\{(\nu,\tau)\}$ to
the set of creators,
and $\{(\nu',\tau')\}$
to the set of annihilators
in the local weight only
(indicated by the union symbol).
This is depicted in
Fig.~\ref{diagramme_monomer1_intro} (b),
with Green's functions
operators
$\hat{c}_\uparrow^\dagger (\tau_2)$
and
$\hat{c}_\uparrow (\tau_2')$ added
to the configuration (c).
Sampling this expansion is referred to as worm sampling or $G$-sampling \cite{wormprb}.

(ii) The standard way to measure the Green's function
($Z$-sampling)
is instead a form of resampling.
The Markov chain produces
partition function configurations
with weight
$w_{\mn{loc}}(\mc{C},\mc{C}') w_{\mn{bath}} (\mc{C},\mc{C}')$,
from which
Green's function configurations are created via
\begin{align}
  G_{\alpha\alpha'}(\tau) &= \frac{1}{Z} \sum_{\mc{C}}
  w_{\mn{loc}}(\mc{C},\mc{C}') w_{\mn{bath}} (\mc{C},\mc{C}')\nonumber \\
  &\times \sum_{n,m=1}^k
  \frac{w_{\mn{bath}} (\mc{C} \backslash \{ (\alpha_n,\tau_n) \}, \mc{C}' \backslash \{ (\alpha_m',\tau_m') \} )}
  %\frac{w_{\mn{bath}} (\mc{C} \backslash \{ \hat{c}_{\alpha_n}(\tau_n) \hat{c}_{\alpha_{n'}}^\dagger(\tau_{n'}) \} )}
  {w_{\mn{bath}} (\mc{C},\mc{C}')} \nonumber \\
  &\times \delta^-\ek{\tau- (\tau_n-\tau_{m}')} \delta_{\alpha\alpha_n} \delta_{\alpha'\alpha_{m}'}.
   \label{eq_z_expansion}
\end{align}
The object after the sum $\sum_{nm}$
is the estimated quantity,
where the original bath weight $w_{\mn{bath}}(\mc{C},\mc{C}')$ is replaced
by a bath weight,
where
$\{(\alpha_n,\tau_n)\}$
has been removed from the creator vertices
and
$\{(\alpha_m',\tau_m')\}$
from the annihilator vertices.
We indicate this by the set difference.
The antiperiodic Dirac comb is defined by
$\delta \ek{ \tau} = \sum_{ n\in\mathbb{Z} } (-1)^n
\delta\nk{\tau - n\beta}$.
This removal leaves behind two operators in the local weight
without hybridization lines,
making them the Green's function operators.
To exploit the full information of a $Z$-configuration,
the procedure of removing the hybridization lines is applied to all possible pairs
of annihilation and creation operators
by the sum $\sum_{nm}$.
The Green's function configuration in Fig.~\ref{diagramme_monomer1_intro} (b)
was constructed by removing
the hybridization lines of two operators
$\hat{c}_\uparrow^\dagger (\tau_2)$
and
$\hat{c}_\uparrow (\tau_2')$
compared to (a).
This also means,
that to each $G$-configuration
uniquely belongs a $Z$-configuration.

\begin{figure}[tp]
  \centering
    \includegraphics[width=0.7\linewidth]{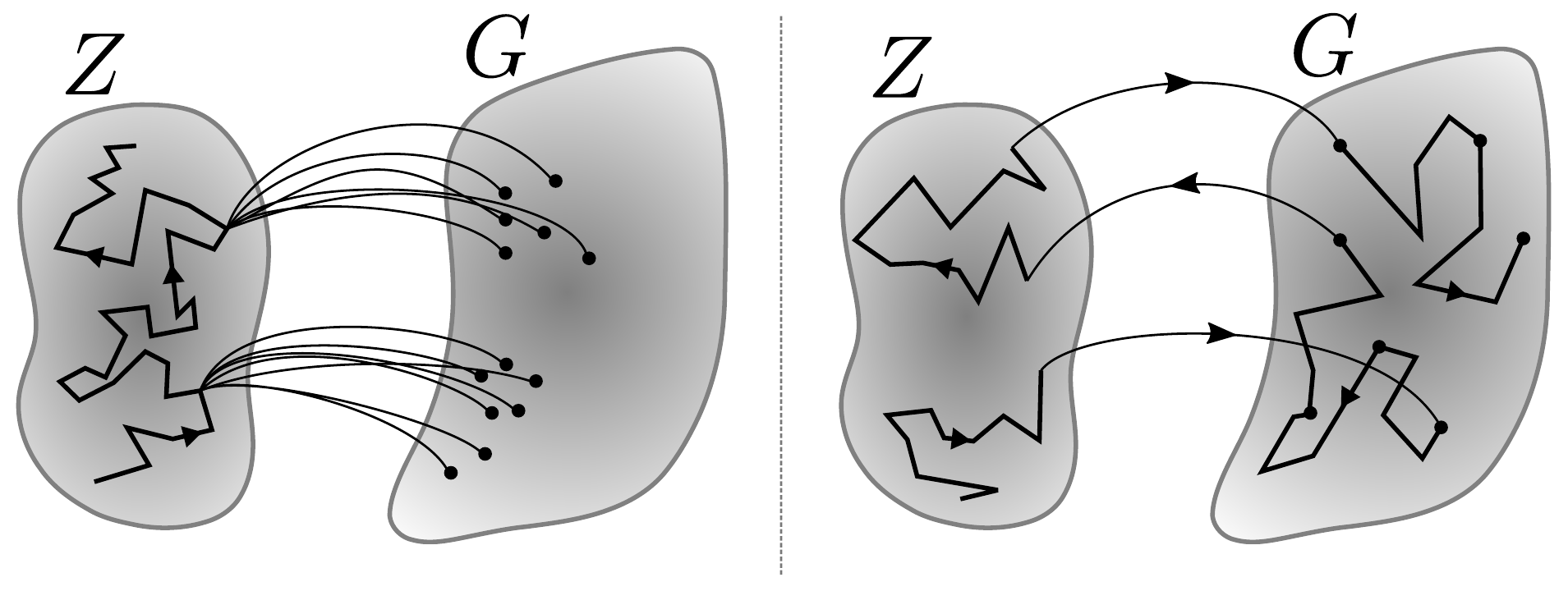}
    \caption{
      Illustration of the two sampling procedures to measure the Green's function.
      Arrows indicate the direction that the random walker moves,
      black circles give the points in $G$-space where the Green's function is measured.
      (ii) In $Z$-sampling (left) the sampling only occurs in $Z$-space, and for generating
      Green's function configurations the $Z$-configuration is changed
      (two hybridization lines are removed).
      (i) In $G$-sampling (right) also $G$-space is sampled.
 }
  \label{worm_and_z}
\end{figure}

Finally, let us illustrate the two sampling procedures
in Fig.~\ref{worm_and_z}.
In $Z$-sampling the sampling solely occurs in $Z$-space;
for the measurement, Green's function configurations are created from $Z$-configurations.
In $G$-sampling the random walk moves between $Z$-space and $G$-space
and samples and measures in both.

\section{Incompleteness of the resampling $G$-estimator}
\label{s_incos}
Here we work out the detected incompleteness
of the resampling estimator for the Green's function.
Equation~\eqref{eq_z_expansion} looks already dangerous:
suppose $w_{\mn{bath}} (\mc{C},\mc{C}')$ is zero,
then the $Z$-configuration $w_{\mn{loc}} (\mc{C},\mc{C}')w_{\mn{bath}} (\mc{C},\mc{C}')$
has zero weight and will never be reached.
Second, if removing the hybridization lines of two operators
gives a nonzero weight
\begin{equation}
w_{\mn{bath}} (\mc{C} \backslash \{( \alpha_n,\tau_n )\}, \mc{C}' \backslash \{( \alpha_m',\tau_m' )\} )
\end{equation}
for a specific pair of $n$ and $m$,
these Green's function configurations will not be generated by $Z$-sampling.
It is known that this is the case
in the atomic limit \cite{wormprb},
as well as for correlators with equal-time operators,
which never occur when sampling the times continuously.

Here we show that it also
happens in systems with a finite number of non-degenerate bath sites,
where the bath
can only host a
finite number of electrons
due to Pauli's exclusion principle.
Suppose a $Z$-configuration deposits one electron more in the bath than the bath can hold.
Then the weight of that $Z$-configuration becomes exactly zero,
but derived $G$-configurations,
which depose one electron less in the bath and thus can have non-zero weight,
are ``missed''.
In the following we call these $Z$-diagrams ``\emph{critical}'', since they would be necessary for obtaining all proper
$G$-configurations.
The AIMs and diagrams
affected by this issue
can be
exactly characterized,
which we will do in the following.
Two ideas of curing the problem are discussed,
namely adding offdiagonal hybridizations
and adding more bath sites,
before we formulate the general criterion to recognize models,
for which the resampling $G$-estimator is incomplete.

\begin{figure}[tp]
  \centering
    \includegraphics[width=1.0\linewidth]{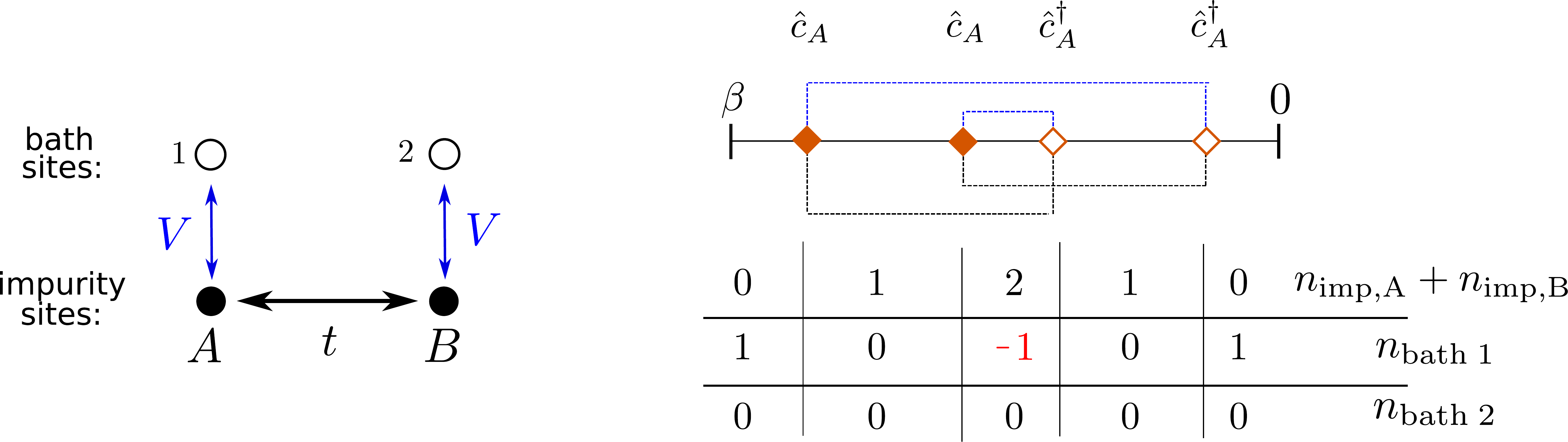}
    \caption{
      Left: sketch of an impurity with two flavors
      $A$ and $B$ (black circles),
      which are connected via a hopping $t$.
      Each of them has a single bath site
      (empty circle),
      to which it couples with amplitude $V$.\\
      Top right: a critical $Z$-configuration of impurity operators
      (the dashed lines indicate
      operators connected to the bath).\\
      Bottom right: a table, which shows the total impurity occupation
      $n_{\mn{imp,A}} + n_{\mn{imp,B}}$ and bath occupations
      $n_{\mn{bath 1}}$ and $n_{\mn{bath 2}}$ for the $Z$-configuration depicted in the middle as a function of imaginary time.
 }
  \label{fig_1bath_diag}
\end{figure}

\subsection{The simplest model: one bath site per impurity site}
\label{sec_one_bath_site}
In Figure~\ref{fig_1bath_diag} we investigate the arguably smallest system
showing the incompleteness:
an impurity \emph{cluster}
of size $N=2$
(two flavors $A$ and $B$, drawn by black circles),
connected by
a single-particle hopping $t$.
Impurity flavor $A$ has a single bath site with label $1$, 
and flavor $B$ one with label $2$
(drawn by empty circles),
to which they couple with amplitude $V$.

Figure~\ref{fig_1bath_diag} also shows
an example $Z$-configuration,
thus all the impurity operators are connected to the bath.
We analyze the occupations of impurity and bath sites.
Reading from right to left,
two electrons of flavor $A$ are created,
which raises the impurity occupation $n_{\mn{imp}} = n_{\mn{imp,A}} + n_{\mn{imp,B}}$
from zero to two.
This is possible,
since the first operator creates the single occupied state
$\ket{A} = \hat{c}^\dagger_A \ket{0}$,
which the time evolution
delocalizes via
$\mn{e}^{\hat{H}\tau} \ket{A} = \alpha \ket{A} + \alpha'\ket{B}$, with $\alpha, \alpha' \in \mathbb{C}$,
due to the single particle hopping $t$
(even though the occupation number basis is
not favorable for implementation,
we use this basis for the discussion here).
Therefore the application of the second operator can give
a doubly occupied impurity state
$\ket{AB} = \hat{c}^\dagger_A \ket{A} + \hat{c}^\dagger_A \ket{B} $
with nonzero amplitude.
The same way two electrons of flavor $A$ can be
annihilated, decreasing the impurity occupation from $2$ to $0$.

However,
there is now no possible way for bath site 1 to receive the two electrons
of flavor $A$.
One either had to end at $\tau=\beta$ with two electrons in bath 1,
which would violate Pauli's principle;
or allow an occupation of -1 as indicated in red in Fig.~\ref{fig_1bath_diag},
which is not possible either.
This makes the weight of this $Z$ configuration zero.
In App.~\ref{app_fk} we discuss in more detail,
how Pauli's principle is implemented in
the language of effective propagators.

If we now create a $G$-configuration
out of the $Z$-diagram in Fig.~\ref{fig_1bath_diag},
we obtain a valid non-zero $G$-diagram
with only one $\hat{c}_A^\dagger$ and $\hat{c}_A$
connected to the bath.
Since it has to be accessed via the zero-weight
$Z$-diagram, it is missed in $Z$-sampling
for this discrete model.
Let us stress
that this $G$-diagram cannot be created
out of another $Z$-diagram,
since in $Z$-sampling every $G$-diagram uniquely belongs
to a $Z$-diagram.

\begin{figure}[tp]
  \centering
    \includegraphics[width=1.0\linewidth]{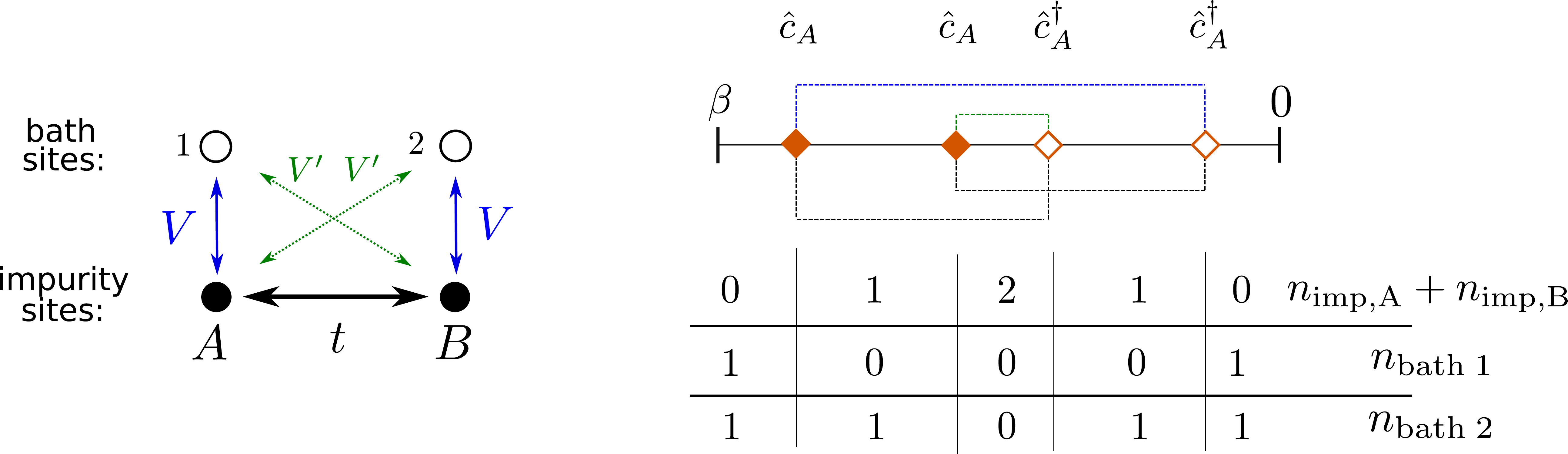}
    \caption{
      Same as Fig.~\ref{fig_1bath_diag} but now with an additional off-diagonal hybridizations $V'$. In this case,
      the Pauli violation of Fig.~\ref{fig_1bath_diag}
      is avoided.
 }
  \label{fig_baths_offdiag}
\end{figure}

\subsection{Adding offdiagonal hybridizations}
Adding off-diagonal hybridization
removes the problem,
since the second electron
of impurity flavor $A$
can be delivered by bath site number $2$
(see Fig.~\ref{fig_baths_offdiag}).
Therefore, critical diagrams
do not exist in clusters with off-diagonal hybridizations,
that have more or the same number of bath sites
compared to the cluster.

\subsection{Adding more bath sites}
One might think
adding more bath sites to the system in Fig.~\ref{fig_1bath_diag}
resolves the problem of missing $G$-diagrams;
this however is only partially true, as we show in this section.

\begin{figure}[tp]
  \centering
    \includegraphics[width=1.0\linewidth]{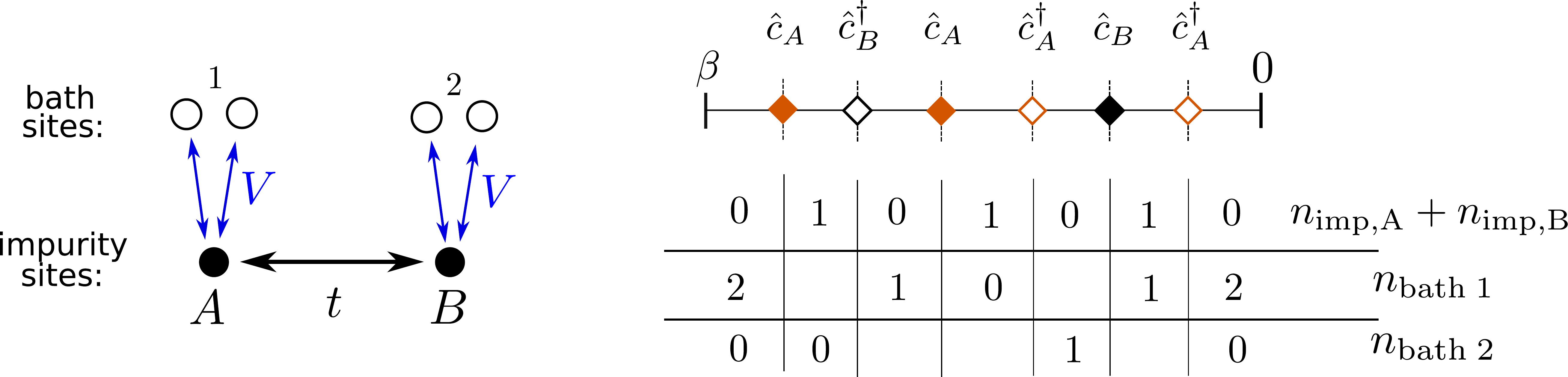}
    \caption{
        A non-critical $Z$-diagram for a system
        with 2 bath sites per impurity flavor.
        For clarity the dashed lines symbolizing
        propagations through the bath have been omitted here.
 }
  \label{fig_Nbaths_diag1}
\end{figure}

In Fig.~\ref{fig_Nbaths_diag1} we see,
that by adding an additional bath site per impurity flavor,
the diagram from the previous Section~\ref{sec_one_bath_site}
is not critical any more.
For reasons that will become clear immediately,
we also added to the trace two operators of flavor $B$.

\begin{figure}[!htp]
  \centering
    \includegraphics[width=0.75\linewidth]{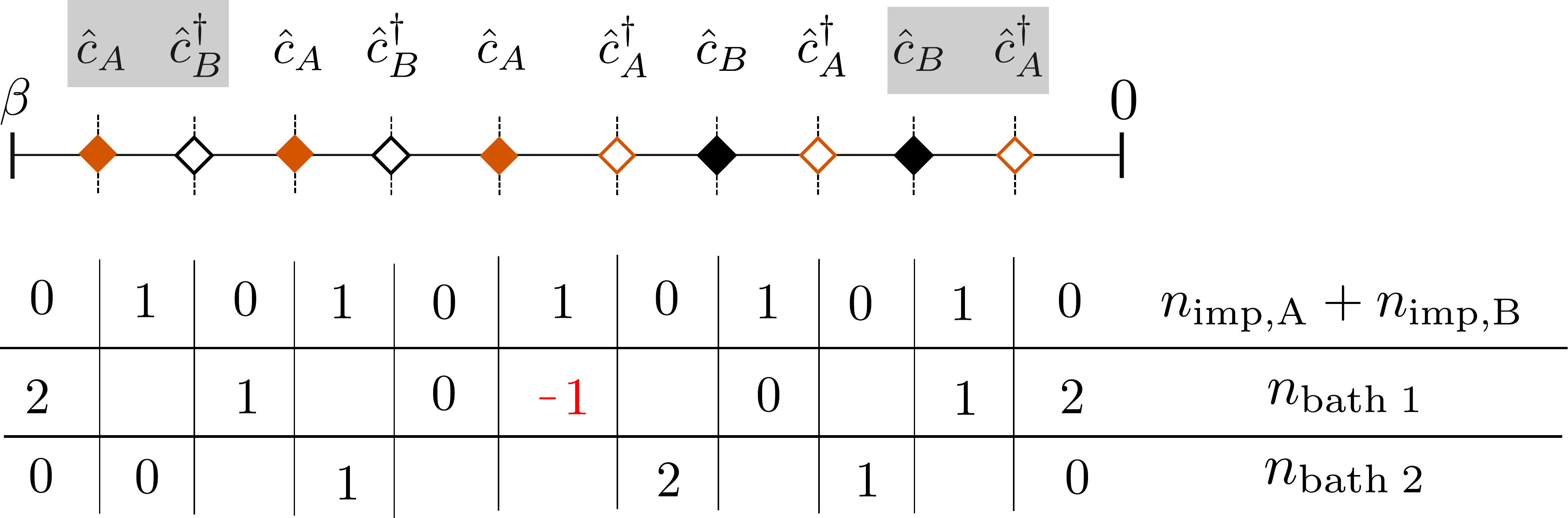}
    \caption{
        A critical diagram for a system
        with 2 non-degenerate bath sites per impurity flavor.
 }
  \label{fig_Nbaths_diag2}
\end{figure}

However,
for this system a critical diagram can still be constructed
by adding a $\hat{c}_A \hat{c}_B^\dagger$ on the left,
and a $\hat{c}_B \hat{c}_A^\dagger$ on the right
(gray boxes in Fig.~\ref{fig_Nbaths_diag2}).
This leads to a construction principle for critical $Z$-diagrams
for this system with $N_\mn{baths}$ non-degenerate bath sites per impurity flavor:
start with $\hat{c}_A \hat{c}_A^\dagger$ and add
$N_\mn{baths}$ times a block of $\hat{c}_A \hat{c}_B^\dagger$ on the left,
and $N_\mn{baths}$ times a block of $\hat{c}_B \hat{c}_A^\dagger$ on the right.

For an impurity cluster, where one impurity site has $N_\mn{baths}$ non-degenerate bath sites,
a $Z$-diagram needs $N_\mn{baths}+1$
impurity creators (annihilators) in a row to be critical,
without annihilators (creators) of the same flavor in between.
However, if $N_\mn{baths}+1$ is larger than the cluster size,
the local trace is not necessarily zero
for exceeding the maximal (subceeding zero) occupation,
since this can get compensated by impurity operators
of the other flavor.

Combinatorics suggest
that for systems with a large number of non-degenerate bath sites,
critical diagrams are much more rare compared to
systems with only a few non-degenerate bath sites;
we will show this explicitly
in Section~\ref{s_monomer1}.

\subsection{General criterion for incomplete resampling $G$-estimator}
In summary, the following conditions suffice
for critical diagrams to exist:
The impurity has a cluster (impurity sites connected by one- or two-particle interaction terms)
of size greater or equal to $2$;
one of the cluster sites has a finite bath,
and there is no hybridization of this finite bath to other flavors.

\section{Numerical analysis}
\label{s_monomer1}
%
% former model taken from /Users/andi/SUPERMUC/benchmarks/1orb_77___beta5___U0.0
% now taken it from /Users/andi/SUPERMUC/benchmarks/1orb_1680___04
%
In this section we provide numerical evidence
for the considerations
formulated in the Section~\ref{s_incos}. For the codes employed, see Appendix~\ref{sec:code}.
First we show
that $Z$-sampling gives wrong Green's functions
for the system in
Fig.~\ref{fig_1bath_diag}.
Then we demonstrate
that critical diagrams are much less important
in systems with more non-degenerate bath sites.
Finally we interpret this outcome by analyzing the autocorrelation times
of critical diagrams.

\subsection{The simplest model}
\label{s_monomer1_general}
The Hamiltonian of the system under consideration
reads
\begin{align}
    \hat{H} &= \begin{pmatrix} \hat{c}^\dagger_\uparrow & \hat{c}^\dagger_\downarrow & \hat{a}_\uparrow^\dagger & \hat{a}_\downarrow^\dagger \end{pmatrix}
        \begin{pmatrix} E_\uparrow & t & V & 0 \\
            t & E_\downarrow  & 0 & V \\
                  V & 0 & \epsilon_\uparrow & 0 \\
                  0 & V & 0 & \epsilon_\downarrow
  \end{pmatrix}
    \begin{pmatrix} \hat{c}_\uparrow \\ \hat{c}_\downarrow \\ \hat{a}_\uparrow \\ \hat{a}_\downarrow \end{pmatrix}
    \label{monomer1_formula}
  \\
    &= \sum_\sigma \left[ E_\sigma \hat{n}_\sigma
    + t \hat{c}^\dagger_\sigma \hat{c}_{\bar{\sigma}}
  + \epsilon_\sigma \hat{a}_\sigma^\dagger a_\sigma
    + V ( \hat{c}_\sigma^\dagger \hat{a}_\sigma + \hat{a}^\dagger_\sigma \hat{c}_\sigma) \right]. \nonumber
\end{align}
The two impurity flavors $E_\sigma$
could differ by any quantum number like
orbital, spin, or a combination thereof.
In order to deal with one-orbital models,
we choose this quantum number to be the spin.
The aberration
of the $Z$-estimator
does not require electron-electron interaction,
hence for simplicity we consider the system to be non-interacting.
Let us note, however, that interactions beyond density-density terms can substitute the role of $t$, as they allow as well
moving electrons from one impurity flavor to another, and thus to
add consecutively a second electron with the same impurity flavor.

The general criterion formulated at the end of Sec.~\ref{s_incos}
is clearly violated here.
We have an impurity cluster of size two,
connected to two single bath sites.

\begin{figure*}[tp]
  \centering
    \includegraphics[width=0.45\columnwidth]{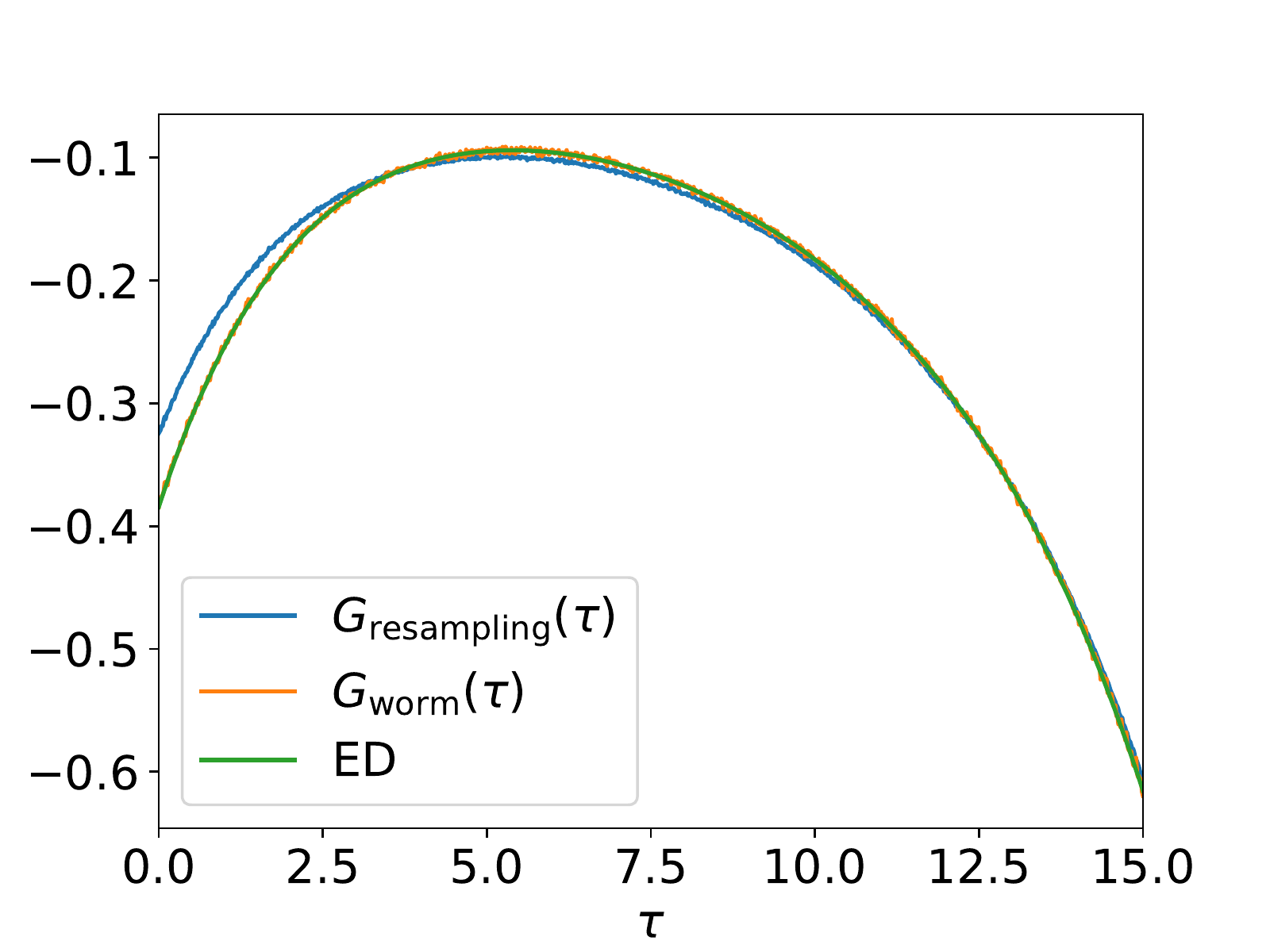}
    \includegraphics[width=0.45\columnwidth]{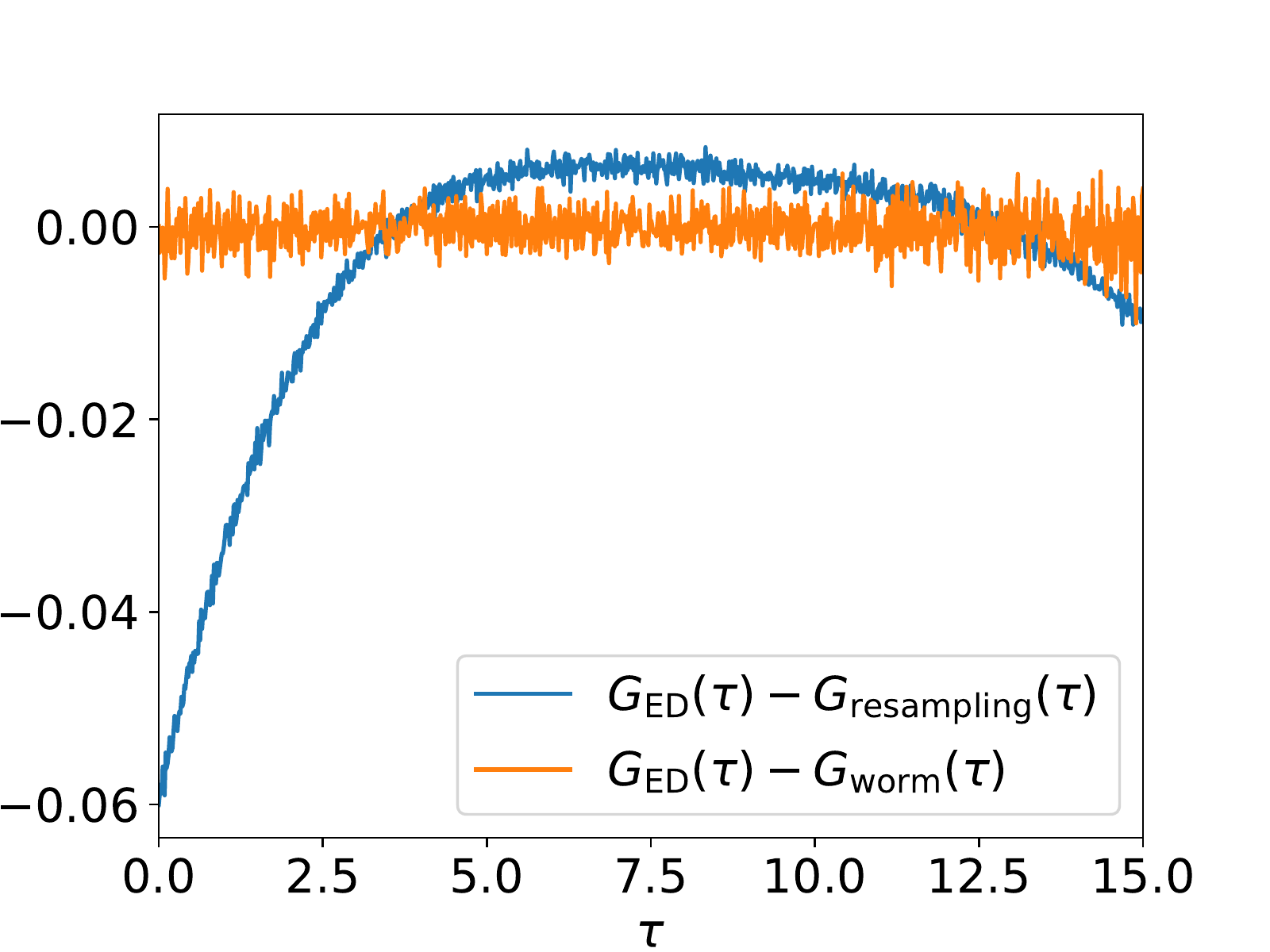}
    \caption{
        Left: Green's function $G_{\uparrow\uparrow}(\tau)$ for model Eq.~\eqref{monomer1_formula}
        with (i) worm sampling and (ii) resampling compared to ED. Right: Differences of the
        QMC results to ED.
 }
  \label{gtau_monomer1}
\end{figure*}

Fig.~\ref{gtau_monomer1} confirms this numerically
for values of
$E_{\uparrow} = E_{\downarrow} = 0.0$,
$\epsilon_\uparrow = 0.2$
$\epsilon_\downarrow = -0.2$,
$t = 0.2$,
$V=0.2$ and an inverse temperature of $\beta = 15.0$.
The Green's functions of $Z$-sampling
are clearly wrong compared to
$G$-sampling or exact diagonalization (ED).
Even the property
\begin{equation}
G_{\sigma\sigma'}(\tau=0^+) + G_{\sigma\sigma'}(\tau = \beta^-) = \delta_{\sigma\sigma'}
  \label{eq_anticomm_rels}
\end{equation}
from the anticommutation relations
of Fermionic operators
is violated in $Z$-sampling,
as one can see from deviation of the
results from resampling compared to ED
in Fig.~\ref{gtau_monomer1}.

\subsection{Autocorrelation time analysis}
\label{ss_autoc}
%%% model taken from: /Users/andi/SUPERMUC/benchmarks/FIND_INFINTE_SYSTEM___02___deltino_0.01
%
We know from Section~\ref{s_monomer1_general},
that resampling for model Eq.~(\ref{monomer1_formula})
misses all
critical diagrams, which have two
annihilation (creation) operators of the same flavor consecutively in a row
(operators of the other flavor may be within that row),
or equivalently, whose bath occupation exceeds (subceeds) 1(0).
Let's decompose the Green's function into two parts
\begin{equation}
  G(\tau) = G_{\mn{crit}}(\tau) + G_{\mn{noncrit}}(\tau),
\end{equation}
where $G_{\mn{crit}}(\tau)$ contains all 
$G$-diagrams derived from the critical 
$Z$-diagrams
for model Eq.~(\ref{monomer1_formula}).
This way the influence
of the critical diagrams
and their autocorrelation times
can be analyzed.

\begin{figure}[tb]
  \centering
    \includegraphics[width=0.7\linewidth]{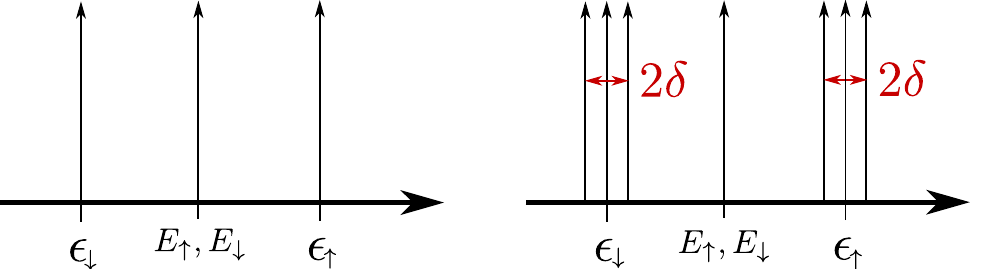}
    \caption{
      Depiction of the discrete energy levels of Hamiltonian \eqref{monomer1_extended_formula}
      for zero and finite $\delta$.
 }
  \label{delta_scan_hamiltonian}
\end{figure}

For this purpose,
we continuously interpolate between a system
with a single bath site per impurity flavor,
and one with three non-degenerate bath sites per impurity flavor.
This can be achieved
by splitting the single bath site into three
and shifting their energy levels apart by a parameter $\delta$ (cf.\ Fig.~\ref{delta_scan_hamiltonian}):
\begin{align}
    \hat{H}'[\delta] &= \begin{pmatrix} \hat{c}^\dagger_\uparrow \\ \hat{c}^\dagger_\downarrow \\ \hat{a}_{1\uparrow}^\dagger \\ \hat{a}_{2\uparrow}^\dagger \\ \hat{a}_{3\uparrow}^\dagger \\ \hat{a}_{1\downarrow}^\dagger \\ \hat{a}_{2\downarrow}^\dagger \\ \hat{a}_{3\downarrow}^\dagger \end{pmatrix} ^\mathrm{T}
        \hspace{0.1cm} \cdot \hspace{0.2cm} \begin{pmatrix} E_\uparrow & t & v & v & v & 0 & 0 & 0 \\
      t & E_\downarrow  & 0 & 0 & 0 & v & v & v \\
                  v & 0 & \epsilon_\uparrow-\delta & 0 & 0 & 0 & 0 & 0\\
                  v & 0 & 0 & \epsilon_\uparrow  & 0 & 0 & 0 & 0\\
                  v & 0 & 0 & 0 & \epsilon_\uparrow + \delta & 0 & 0 & 0 \\
                  0 & v & 0 & 0 & 0 & \epsilon-_\downarrow\delta & 0 & 0\\
                  0 & v & 0 & 0 & 0 & 0 & \epsilon _\downarrow & 0 \\
                  0 & v & 0 & 0 & 0 & 0 & 0 & \epsilon_\downarrow + \delta
  \end{pmatrix}
  \hspace{0.2cm}\cdot\hspace{0.2cm}
    \begin{pmatrix} \hat{c}_\uparrow \\ \hat{c}_\downarrow \\ \hat{a}_{1\uparrow} \\ \hat{a}_{2\uparrow} \\ \hat{a}_{3\uparrow}  \\ \hat{a}_{1\downarrow} \\ \hat{a}_{2\downarrow} \\ \hat{a}_{3\downarrow}
  \end{pmatrix}.
    \label{monomer1_extended_formula}
\end{align}
We further discuss this construction
in appendix \ref{app_dummy_sites}.
For consistency with the previous model Eq.~\eqref{monomer1_formula} the
new hybridization is $v=V/\sqrt{3}$.
One can verify by using the resolvent expression $G = (\mathbb{1}i\omega -\hat{H})^{-1}$ or calculating
the hybridization function, that the systems 
$\hat{H}$ and $\hat{H}'[\delta=0]$
are identical.

\begin{figure*}[!htpb]
  \centering
    \includegraphics[width=1.0\linewidth]{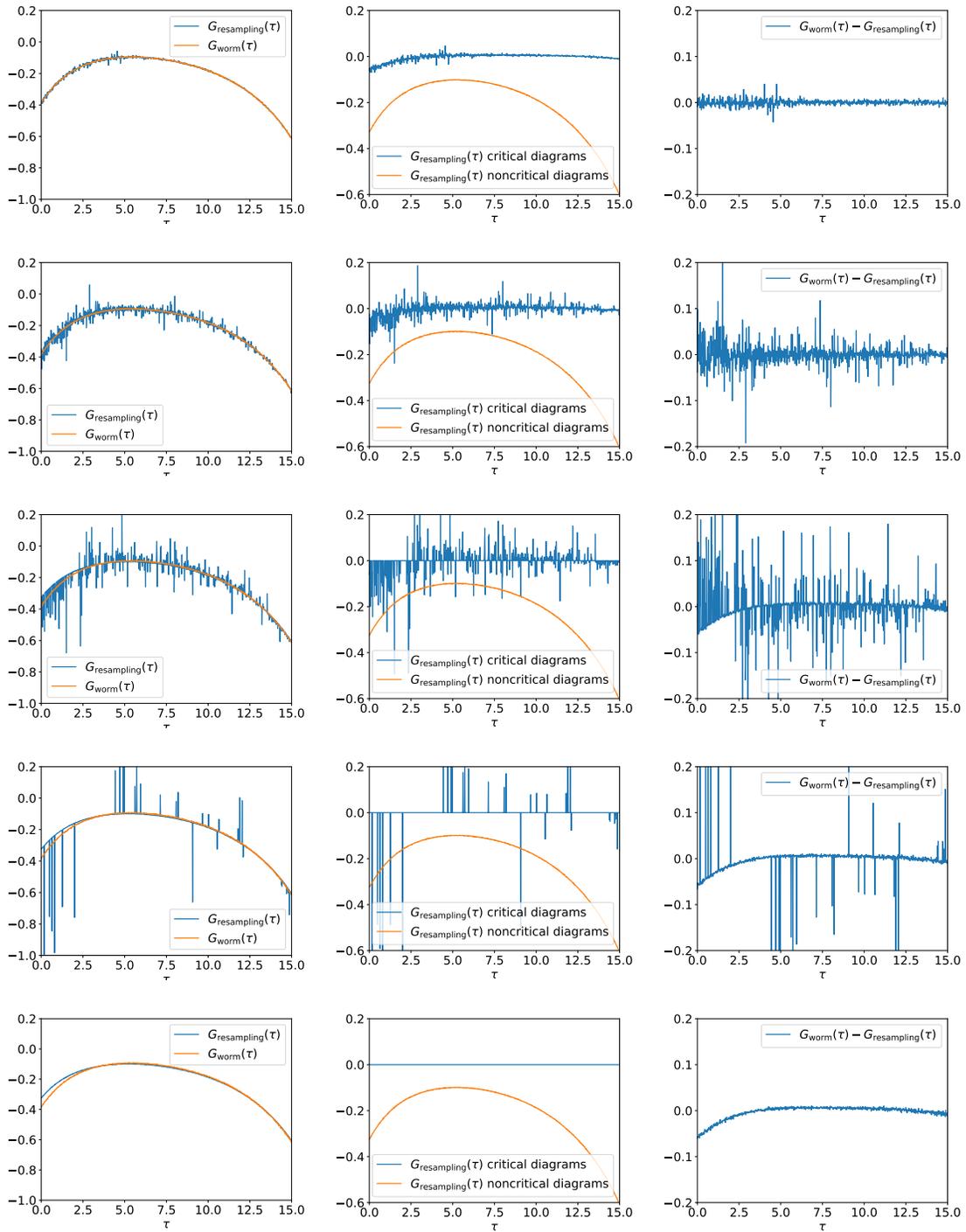}
    \caption{
        From top to bottom: Hamiltonian $H'[\delta]$ of Eq.~(\ref{monomer1_extended_formula})
        for $\delta$ = 0.05, 0.01, 0.003, 0.001, 0.0.
      The left column shows the Green's functions for resampling and worm sampling,
      the middle column a distribution of the resampling Green's function
      in critical and noncritical diagrams,
      and the right column shows the difference between
      worm and resampling results.
 }
  \label{delta_scan}
\end{figure*}

A scan over the parameter $\delta$ is shown in Fig.~\ref{delta_scan}
for $\delta \in \{ 0.05, 0.01, 0.003, 0.001, 0.0 \} $.
Since the number of Monte Carlo steps and measurements
is the same for all panels
and the systems are very similar,
the difference in the noise can be regarded as
an autocorrelation effect.
$Z$-sampling agrees with ED for
the system with $\delta=0.05$;
this system can be considered as having three non-degenerate bath sites per impurity flavor,
therefore the contributions of
critical diagrams for $N_\mn{baths per flavor}=1$
are all produced frequently in $Z$-sampling.
The critical diagrams for
$N_\mn{baths per flavor}=3$
are not produced, but their absence is clearly
not visible with
the Monte Carlo precision of this calculation.
Upon decreasing $\delta$
the noise increases strongly,
because $Z$-sampling is less and less able to produce the
critical diagrams for $N_\mn{baths per flavor}=1$.
For $\delta=0.00$ none of those
can be produced by the $Z$-sampling any more.
Their contribution is zero without noise
and the result is wrong.
In other words, upon approaching the $N_\mn{baths per flavor}=1$
limit, the autocorrelation times of the
critical diagrams increases, and becomes infinite
when reaching the limit.
The curve for the noncritical diagrams is smooth all the time,
which means their autocorrelation times are small and independent of $\delta$.
Worm sampling is instead able to produce all
$G$-diagrams without problems.

The same observation can be made if we --instead of $\delta$-- extend the Hamiltonian of Eq.~(\ref{monomer1_formula})
by an offdiagonal hybridization $V'$:
\begin{align}
    \hat{H}''[V'] &= \begin{pmatrix} \hat{c}^\dagger_\uparrow & \hat{c}^\dagger_\downarrow & \hat{a}_\uparrow^\dagger & \hat{a}_\downarrow^\dagger \end{pmatrix}
        \begin{pmatrix} E_\uparrow & t & V & V' \\
            t & E_\downarrow  & V' & V \\
                  V & V' & \epsilon_\uparrow & 0 \\
                  V' & V & 0 & \epsilon_\downarrow
  \end{pmatrix}
    \begin{pmatrix} \hat{c}_\uparrow \\ \hat{c}_\downarrow \\ \hat{a}_\uparrow \\ \hat{a}_\downarrow \end{pmatrix}
    \label{hamilt_offd_h}
\end{align}
\begin{figure*}[!htpb]
  \centering
    \includegraphics[width=\linewidth]{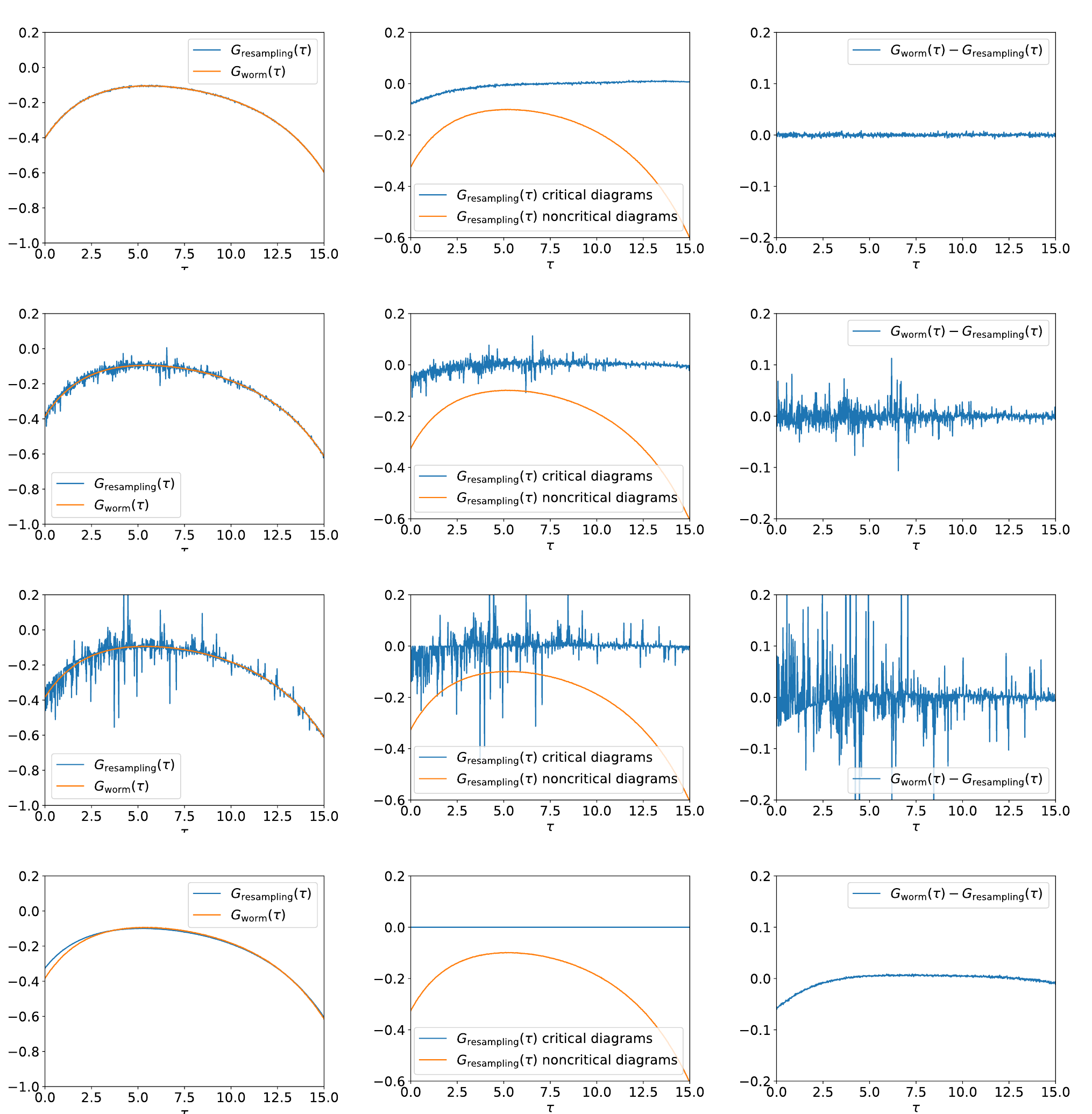}
    \caption{
      From top to bottom: Hamiltonian $H''[V']$ of Eq.~(\ref{monomer1_extended_formula})
      for $V'$ = 0.03, 0.003, 0.001, 0.0.
      The left column shows the Green's functions for resampling and worm sampling,
      the middle column a distribution of the resampling Green's function
      in critical and noncritical diagrams,
      and the right column shows the difference between
      worm and resampling results.
 }
  \label{offd_scan}
\end{figure*}
Figure~\ref{offd_scan} shows,
that for $V'\neq 0$ resampling
agrees with G-sampling
(in this case, gives the correct result),
for small $V'$ it gives noisy but correct results,
and for $V'=0$ it gives the wrong result.

\section{Continuous baths}
\label{s_inf_sys}
%
%%% taken from FIND_INFINTE_SYSTEM___05___2_baths_far_away___weak_coupling___notsofaraway_6___make_infinite___6
%
Systems with an infinite number of non-degenerate bath sites do not
produce incomplete estimators of the type discussed in Section~\ref{s_incos}.
However, resampling may still suffer from
autocorrelation problems,
if a part of the bath is well-approximated by a discrete one,
as we will show now.

For this purpose we take an AIM
with a bath consisting of two parts:
a finite part
(which on its own had an incomplete resampling $G$-estimator)
and an infinite one,
such that for the whole system resampling for $G$ is surjective.
Its Hamiltonian is
\begin{align}
  \hat{H} =& \hat{H}_{\mn{imp}}
  + \hat{H}_{\mn{disc. bath \& hyb.}}
  + \hat{H}_{\mn{cont. bath \& hyb.}} 
    \label{eq_inf_sys} \\
  =& \sum_\sigma (E_\sigma \hat{c}^\dagger_\sigma \hat{c}_\sigma + t \hat{c}^\dagger_\sigma \hat{c}_{\bar{\sigma}} ) \nonumber  \\
  &+ \sum_\sigma \epsilon_\sigma \hat{a}_\sigma^\dagger \hat{a}_\sigma
   + \sum_\sigma V ( \hat{c}^\dagger_\sigma \hat{a}_\sigma + \hat{a}_\sigma^\dagger \hat{c}_\sigma) \nonumber  \\
  &+ \hat{H}_{\mn{cont. bath \& hyb.}}
\end{align}
\begin{figure}[!htp]
  \centering
  \includegraphics[width=0.45\linewidth]{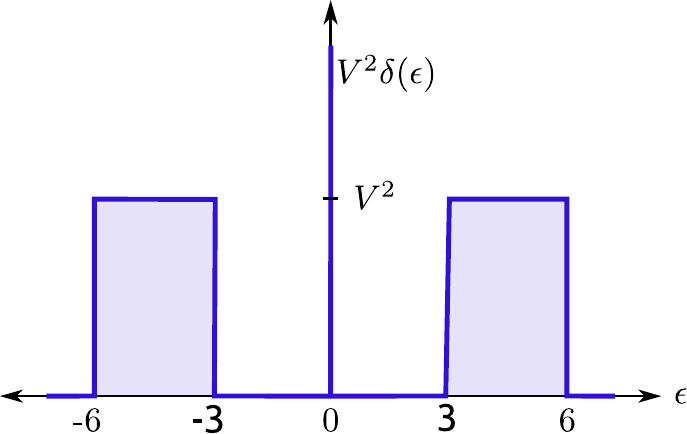}
    \caption{
Hybridization function considered for autocorrelation analysis
in systems with infinite baths.
It comprises of a delta-peak at $\epsilon=0$ and higher-energy bands.
This system is used to mimic a situation with a very narrow peak at the
Fermi level and extended hybridization at higher energies.\\
 }
  \label{inf_system_ftau}
\end{figure}
The impurity has again two flavors $\sigma\in\{\uparrow, \downarrow\}$
with respective energy levels $E_\sigma$,
connected by a single-particle hopping $t$.
Each impurity flavor is connected to a single bath level with energy
$\epsilon_{\sigma}$ and amplitude $V$.
For the system $H_{\mn{imp}} + H_{\mn{disc. bath \& hyb.}}$ alone,
critical $Z$-diagrams
caused problems with $Z$-sampling,
as shown in Sec.~\ref{s_monomer1_general}.
The Hamiltonian extended
with a continuous bath
$H_{\mn{cont. bath \& hyb.}}$,
specifically bands at higher energy,
is not supposed to show an incomplete resampling $G$-estimator.
Its hybridization function on the real frequency axis is shown in Fig.~\ref{inf_system_ftau}.
%
%
%-0.0054*(np.log(iw.value + 5) - np.log(iw.value + 6) ) -0.0054*(np.log(iw.value - 6) - np.log(iw.value - 5)) + 0.3**2/(iw.value-0.1)
%
%
\begin{figure}[!htp]
  \centering
    \includegraphics[width=1.0\linewidth]{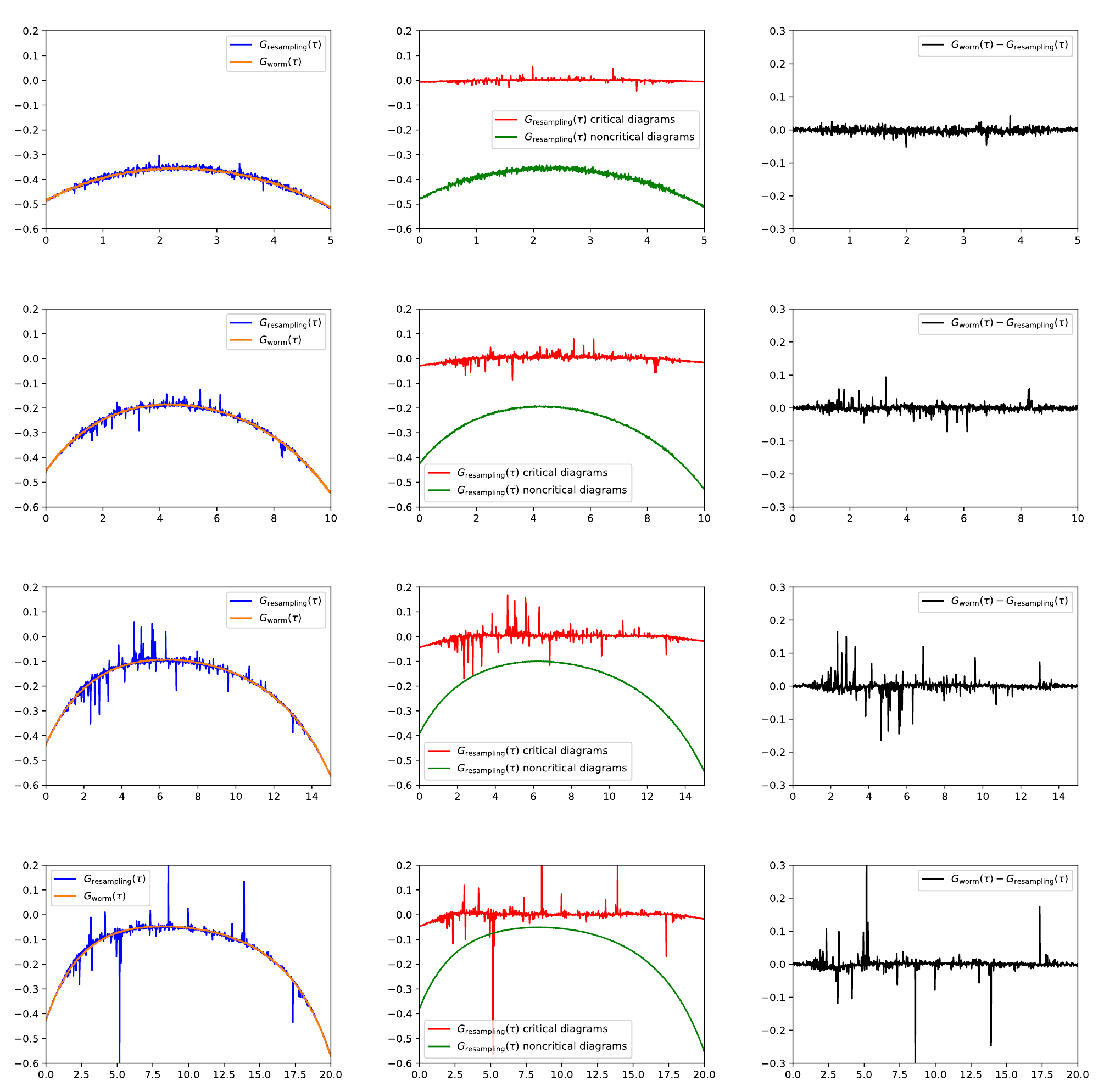}
    \caption{
        Data for Hamiltonian Eq.~\ref{eq_inf_sys}.
From top to bottom: inverse temperatures $\beta = 5$, $10$, $15$ and $20$.
Left column: Greens functions for resampling and worm sampling, middle:
splitting of diagrams of resampling in critical and noncritical
with respect to the finite subsystem; right: deviation between
resampling and worm sampling.
It is clear that the noise of the resampling method essentially stems from
the critical diagrams.
 }
  \label{inf_system_data}
\end{figure}
We transform the hybridization from real frequencies to
Matsubara frequencies via
\begin{align}
  \Delta(i\omega_n) &=  \frac{V^2 }{ i\omega} + \int_{-6}^{-3} \mn{d}\epsilon \; \frac{V^2 }{ i\omega -\epsilon} + \int_{3}^{6} \mn{d}\epsilon \; \frac{V^2 }{ i\omega -\epsilon} .
                      %&= V^2 ( -\mn{ln}(-i\omega_n - 6) - \mn{ln}(-i\omega_n - 5) ) \\
                      %&+ V^2 ( \mn{ln}(-i\omega_n + 5) - \mn{ln}(-i\omega_n + 6) ) \\
                      %&+ \frac{V^2}{i\omega_n }.
\end{align}
For the numerical calculation
we set the hybridization strength
to $V=0.3$, $\epsilon_{\uparrow}=0.1$, $\epsilon_{\downarrow}=-0.1$,
and especially the hopping $t=0.2$ to a physically more realistic value.

Let us now investigate the autocorrelation times for this system.
As it can be seen in Fig.~\ref{inf_system_data},
the worm-sampling produces precise data for 
all considered inverse temperatures of $\beta = 5$, $10$, $15$ and $20$.
We now separate the resampling $G(\tau)$ into two parts,
a critical and a non-critical one, with respect to
the system $H_{\mn{imp}} + H_{\mn{disc. bath \& hyb.}}$.
For $\beta=5$ both show the same amount of noise.
Upon lowering the temperature,
the non-critical part gets smoother,
which is to be expected, since the expansion order
grows linearly with $\beta$, therefore more information can be extracted from a single diagram.
However, the larger expansion order gives the Monte Carlo combinatorically more options
to suffer Pauli violations in the bath with respect to the discrete part of $\hat{H}$;
this forces the electrons now to propagate through the high energy satellites,
and therefore damps the weight of the corresponding $Z$-diagrams significantly.
This leads to increased autocorrelation time of the corresponding critical $G$-diagrams,
as can be seen in the noise.
The results for $Z$-sampling seem to be correct for $\beta = 5$ and $10$
and fulfill the sum rule in Eq.~(\ref{eq_anticomm_rels}).
The results for $\beta=15$ and $20$ are not converged for $Z$-sampling
with this statistics.

This example proves, that the phenomenon described in this work
can also significantly affect models with an infinite number
of non-degenerate bath sites.

\section{Conclusion}
\label{sec:conclusion}
We found that the standard CTHYB estimator of
the Green's function ($Z$-sampling) unexpectedly fails for the
Anderson impurity model in some, hitherto unknown, cases. Specifically,
this  aberration occurs for clusters with $N\geq 2$ flavors on the impurity,
where at least one cluster site couples to a discrete, finite bath
and this bath does not couple to other impurity flavors through offdiagonal hybridizations.
Pauli's exclusion principle forces the weight
of some partition function configurations to be zero.
This is problematic since nonzero Green's function configurations
would need to be generated out of these
unreachable partition function configurations.
In worm sampling ($G$-sampling), this
kind of 
aberration
does not occur.

Furthermore our findings explain the
occurrence of large autocorrelation times
for systems, whose infinite baths can well
be approximated by finite baths,
and limit the application of CTHYB
for quantum chemistry applications
to worm sampling only.
In general,
our findings illustrate,
that for any Markov chain Monte Carlo algorithm
of diagrammatic series,
it is very important to carefully ensure
surjectivity of the
mapping between the sampled distribution
and observable distribution.

\section{Acknowledgments}
We thank Olivier Parcollet and especially Nils Wentzell for useful discussions. 
A. H. and G. S. thank the Simons Foundation for the  hospitality at the CCQ of the Flatiron Institute,
which is a division of the Simons Foundation.

%% TODO: include author contributions
%\paragraph{Author contributions}
%This is optional. If desired, contributions should be succinctly described in a single short paragraph, using author initials.

% TODO: include funding information
\paragraph{Funding information}
A. H. and G. S. acknowledge financial support from the DFG through Würzburg-Dresden Cluster of Excellence on Complexity 
and Topology in Quantum  Matter — ct.qmat (EXC 2147, project-id 390858490). 
M. W. and K. H. acknowledge the FWF (Austrian Science Funds) through project P32044.
We further acknowledge funding through the Research Unit "Quast"
funded by the DFG as project FOR-5249 (G.S.; project P4) and the FWF as I5868 (K.H.; project P1).
We are grateful to the Gauss Centre for Supercomputing e.V. (www.gauss-centre.eu) 
for funding this project by providing computing time on the GCS Supercomputer SuperMUC-NG at Leibniz Supercomputing Centre (www.lrz.de).

\begin{appendix}
\section{Software}
\label{sec:code}
The CTHYB data was produced with w2dynamics \cite{w2dyn}
using an interface \cite{w2dyninterface}
to the TRIQS library \cite{triqs}.
The results shown in Fig.~\ref{gtau_monomer1}
were confirmed with the CTHYB solver from TRIQS \cite{triqscthyb}.
The ED calculations were done with pomerol \cite{pomerol},
also using its interface to TRIQS.
\section{Technical remarks}
\label{sec:technical}

Let us note that the problem of incompleteness discussed here is not caused by
an accidental zero of the weight due to numerical noise.
However $Z$-configurations with exactly zero weight
can have nonzero weight due to numerical instabilities,
especially when using Sherman-Morrison formulas
for updating the bath weight
after insertion / removal of diagram vertices.
Then it is possible that the forbidden $G$-configurations
can still be accessed with correct probabilities,
since the ``wrong'' but nonzero weights of the $Z$-configurations
cancel out.
The authors observed such a case,
where in resampling $G(\tau)$ first seemingly converged to a wrong result,
then it became spiky
and showed a very slow convergence
towards the correct result.
This may happen especially for large expansion orders, i.e. small temperatures.
Worm sampling instead immediately converged to the correct result.

\section{Pauli's principle for effective propagators by the example an bath with one site}
\label{app_fk}
Here we will discuss
how Pauli's principle is implemented
for effective propagators
using the $Z$-diagram in Fig.~\ref{fig_1bath_diag} (a)
as an example.
The hybridization function of Hamiltonian \eqref{monomer1_formula} is
\begin{equation}
    \Delta_{\mn{AA}}(\tau) = \frac{V^2}{\mn{e}^{\beta\epsilon}+1} \times
  \begin{cases}
    \mn{e}^{\epsilon \tau}, & \mn{if}\ \tau > 0 \\
   -\mn{e}^{\epsilon (\beta-\tau)}, & \mn{if}\ \tau < 0
  \end{cases},
\end{equation}
where the bath site was integrated out.
Summing the effective propagators to a determinant
for the $Z$-configuration of Fig.~\ref{fig_1bath_diag} (a) gives

\begin{align}
  w_\mn{bath} (\mc{C}, \mc{C}')
  &= \mn{det}
  \begin{pmatrix}
    \Delta_{AA}(\tau_1-\tau_1') & \Delta_{AA}(\tau_1-\tau_2') \\
    \Delta_{AA}(\tau_2-\tau_1') & \Delta_{AA}(\tau_2-\tau_2')
  \end{pmatrix} \\ \nonumber
  &=
    \left( \frac{V^2}{\mn{e}^{\beta\epsilon}+1} \mn{e}^{\epsilon \beta} \right)^2
   \mn{det} \begin{pmatrix}
     \mn{e}^{-\epsilon (\tau_1-\tau_1')} & \mn{e}^{-\epsilon (\tau_1-\tau_2')} \label{eq_rank_def_mat}\\
    \mn{e}^{-\epsilon (\tau_2-\tau_1')} & \mn{e}^{-\epsilon (\tau_2-\tau_2')}
   \end{pmatrix} \nonumber \\
  &= 0,
\end{align}
with $\tau_1<\tau_2<\tau_1'<\tau_2'$.
The matrix in Eq.~\eqref{eq_rank_def_mat} is rank deficient:
the rows are the same,
but the first is multiplied with a factor $\mn{e}^{\epsilon \tau_1'}$,
the second with a
factor $\mn{e}^{-\epsilon \tau_2'}$.
This is Pauli's exclusion principle
in the language of effective propagators.
In the actual CTHYB code this is how, at a certain time, only one
electron is allowed to propagate through the bath.
The generalization to bigger matrices is straightforward.

\section{Energy degenerate bath sites}
\label{app_dummy_sites}
One may think that the violation of the Pauli principle
in a discrete-bath system could be circumvented
by adding bath sites that duplicate the existing ones,
hence hosting the necessary number of electrons
that a given configuration requires.
This situation corresponds to the $\delta=0$ case
in Eq.~\eqref{monomer1_extended_formula}.
The additional bath sites instead
effectively decouple from the impurity as we show in the following by applying a unitary transformation of the
bath degrees of freedom.

Let's for simplicity start with the Hamiltonian operator
of one impurity and bath site
\begin{align}
  \hat{H} &= \begin{pmatrix} \hat{c}^\dagger & \hat{a}_1^\dagger  \end{pmatrix}
  \begin{pmatrix} E & v  \\
    v & \epsilon
  \end{pmatrix}
    \begin{pmatrix} \hat{c} \\ \hat{a}_1 \end{pmatrix}
    \label{app_ham_1}
\end{align}
and duplicate the bath site:
\begin{align}
    \hat{H}' &= \begin{pmatrix} \hat{c}^\dagger & b_1^\dagger & b_2^\dagger  \end{pmatrix}
      \underbrace{  \begin{pmatrix} E & v & v \\
    v & \epsilon  & 0 \\
                  v & 0 & \epsilon
      \end{pmatrix} }_{\mathcal{H}'}
    \begin{pmatrix} \hat{c} \\ \hat{b}_1 \\ \hat{b}_2  \end{pmatrix}.
  \label{app_ham_2}
\end{align}
We can apply the following unitary transformation
\begin{equation}
  A = \begin{pmatrix}
    1 & 0 & 0 \\
    0 & \frac{1}{\sqrt{2}} & \frac{1}{\sqrt{2}}  \\
    0 & \frac{1}{\sqrt{2}} & -\frac{1}{\sqrt{2}}
  \end{pmatrix},
\end{equation}
to the Hamiltonian matrix $\mathcal{H}'$ of $\hat{H}'$ and find
\begin{equation}
    A^\dagger \mathcal{H}' A =
    \begin{pmatrix} E & \sqrt{2}v & 0 \\
    \sqrt{2}v & \epsilon  & 0 \\
                  0 & 0 & \epsilon
  \end{pmatrix}.
  \label{app_ham_3}
\end{equation}
This object can host two electrons in its bath degrees of freedom,
but only one of them is connected to the impurity site
with a rescaled hybridization strength of $\sqrt{2}v$.
Since Eqs.~\eqref{app_ham_2} and \eqref{app_ham_3}
are simply the same objects in other bases,
in Eq.~\eqref{app_ham_2} also only one electron can hop
from the impurity into the bath.
However, splitting the two bath energies by $\delta$ removes
this restriction.

% TODO:
% Provide your bibliography here. You have two options:

% FIRST OPTION - write your entries here directly, following the example below, including Author(s), Title, Journal Ref. with year in parentheses at the end, followed by the DOI number.
%\begin{thebibliography}{99}
%\bibitem{1931_Bethe_ZP_71} H. A. Bethe, {\it Zur Theorie der Metalle. i. Eigenwerte und Eigenfunktionen der linearen Atomkette}, Zeit. f{\"u}r Phys. {\bf 71}, 205 (1931), \doi{10.1007\%2FBF01341708}.
%\bibitem{arXiv:1108.2700} P. Ginsparg, {\it It was twenty years ago today... }, \url{http://arxiv.org/abs/1108.2700}.
%\end{thebibliography}

\bibliographystyle{SciPost_bibstyle} % Include this style file here only if you are not using our template
\bibliography{SciPost_Example_BiBTeX_File}

\end{appendix}

\nolinenumbers

\end{document}